\documentclass[sigconf,anonymous=false]{acmart}
\renewcommand\footnotetextcopyrightpermission[1]{} 
\usepackage{graphicx}
\usepackage{amsmath}
\usepackage{booktabs}
\usepackage{algorithm}
\usepackage{algorithmic}
\usepackage{amsfonts}
\usepackage{multirow}[c]
\usepackage{makecell}
\usepackage{subfigure}
\usepackage{color}
\usepackage{bm}
\usepackage{epstopdf}
\usepackage{url}
\usepackage[cal=cm]{mathalfa}
\usepackage{balance}
\usepackage{threeparttable}
\usepackage{wrapfig}
\usepackage{tabularx}
\usepackage{array}
\usepackage{enumitem}
\usepackage{appendix}
\usepackage{tikz}
\usepackage{pifont}

\setlist[itemize]{leftmargin=*}

\makeatletter
\newcommand\notsotiny{\@setfontsize\notsotiny\@vipt\@viipt}
\makeatother

\AtBeginDocument{%
  }

\begin{document}


\title{LazUnirec: Unifying Retrieval and Ranking in Multi-scenario\\ Industrial Recommendation}

\author{Xiaodong Li, Junfeng Zhang, Shichao Nie, Ziyue Guo, Yiming Liu, Xinxin Xu, Xurong Li,\\ Jinyuan Dang, Zhanming Ou, Jianan Yan*, Tao Luo, Xiaoyi Zeng}
\thanks{* Corresponding author.}
\affiliation{
\institution{Alibaba International Digital Commerce Group, Beijing, China}
\country{\{limingxi.lxd, sichu.zjf, nsc383441, guoziyue.gzy, lym559243, xxx558853, xurong.li, dangjinyuan.djy,\\ zhanming.ozm, jianan.yjn, luotao.lt, yuanhan\}@alibaba-inc.com}
}

\renewcommand{\shorttitle}{LazUnirec: Unifying Retrieval and Ranking in Multi-scenario Industrial Recommendation}
\renewcommand{\shortauthors}{Xiaodong Li et al.}

\begin{abstract}

Modern industrial recommendation systems typically divide retrieval and ranking into two separate cascaded stages, leading to inconsistent optimization objectives, redundant encoding of user sequences, and a waste of computational resources. Previous efforts have investigated using Transformers to model both retrieval and ranking simultaneously.
However, these studies still face two major limitations: (1) Since retrieval and ranking share a single Transformer backbone, the gradient conflict arises when the gradient directions of the two tasks are inconsistent. (2) When gradient conflict occurs, the model cannot determine a reliable direction for gradient updates.
To this end, we propose a \textit{Unified Retrieval and Ranking approach for Multi-scenario Industrial Recommendation}, termed as \textbf{LazUnirec}. 
Specifically, we construct task-shared and task-private parameters on the Transformer backbone to mitigate the gradient conflict.
Wherein, for the attention components, we propose LoRA-modulated attention, enabling retrieval and ranking to share attention parameters while preserving task-private parameters. 
For the feed-forward networks (FFN), we present task-adaptive sparse mixture-of-experts (MoE) to mitigate gradient conflict by constructing task-shared and task-private experts.
Moreover, we propose an uncertainty estimation that evaluates the gradient reliability of the retrieval and ranking, thereby guiding the model toward a more reliable gradient update direction when gradient conflict occurs.
In addition, knowledge distillation from ranking to retrieval further enables cross-task knowledge transfer. 
Extensive offline experiments and online A/B testing demonstrate the effectiveness and superiority of LazUnirec. LazUnirec has been deployed in online multi-scenario industrial recommendation system and currently delivers significant gains for the platform.

\end{abstract}



\maketitle

\begin{figure}[t]
\setlength{\abovecaptionskip}{0.cm}
	\begin{center}
        \subfigure
        {\begin{minipage}[b]{.49\linewidth}
        \centering
        \includegraphics[scale=0.135]{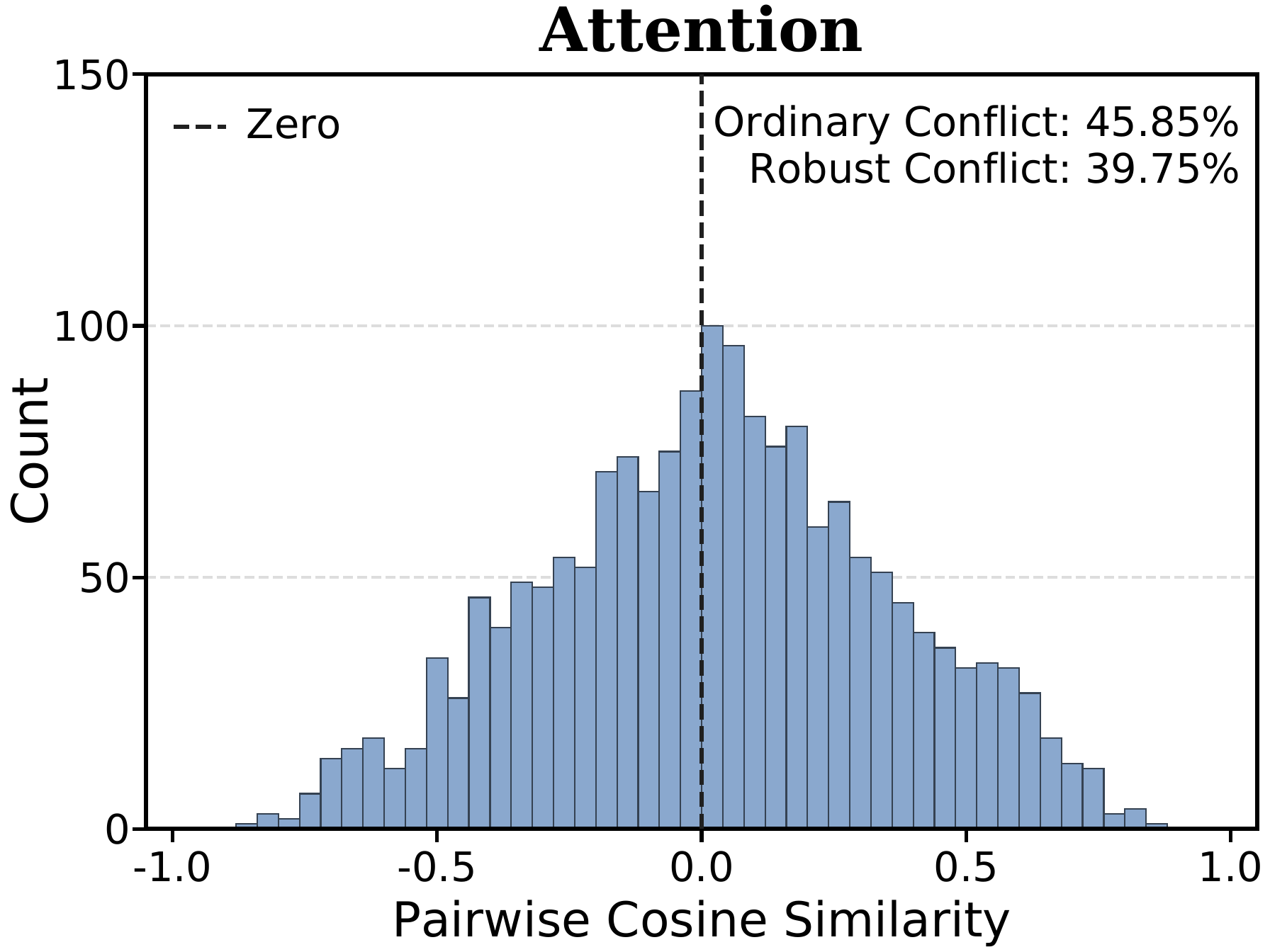}
        \end{minipage}}
        \subfigure
        {\begin{minipage}[b]{.49\linewidth}
        \centering
        \includegraphics[scale=0.135]{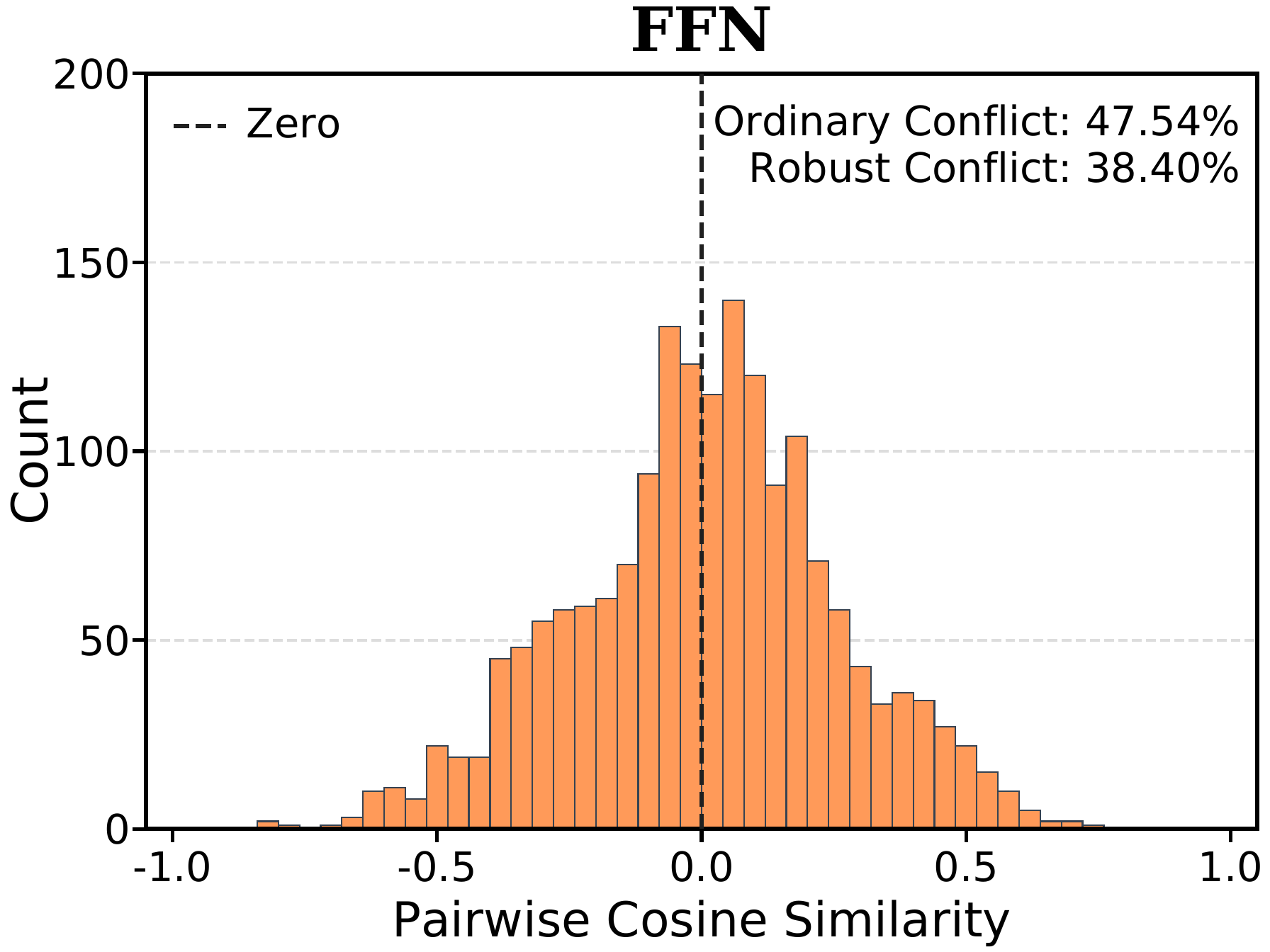}
        \end{minipage}}
        
	\caption{Statistics on the percentage distribution of gradient conflicts for retrieval and ranking within Transformer backbone, including Attention and FFN components.}
	\label{motivation}
	\end{center}
\vspace{-0.3cm}
\end{figure}

\section{Introduction}

Modern industrial recommendation systems typically separate retrieval and ranking into two independently cascaded stages. Retrieval~\cite{dnnyr,gcnnws,facet,sdm} aims to efficiently identify a subset of items that users are likely to interact with from a item corpus, while ranking~\cite{farm,rankmixer,tokenmixer_large} predicts multiple probabilities (\textit{i.e.}, CTR, CVR) to characterize users' precise preferences for candidate items. 
This division has long been the dominant paradigm in the industry recommendation, as it effectively balances efficient search with accurate classification. 
However, these two stages are often encoded based on the same user sequence, while maintaining separate parameters and optimization objectives.
Consequently, this cascaded design introduces inconsistent optimization objectives, redundant encoding of user sequences, and unnecessary consumption of computational resources. As industrial systems process longer user sequences and serve multiple recommendation scenarios, these costs become increasingly significant.

At the same time, both retrieval and ranking are increasingly adopting Transformer~\cite{trans} architectures. On the one hand, retrieval leverage Transformers to encode historical interactions and extract user preferences for candidate generation. Some studies~\cite{onerec_tr,onerec,tiger,onelive,oneloc,onemall} further adopt a generative paradigm, using semantic IDs to represent each item for autoregressive prediction. On the other hand, some ranking methods~\cite{onetrans,lazformer} employ Transformer to jointly conduct user sequence modeling and feature interaction, which benefits from scaling depth, width, and sequence length. Although their outputs and prediction objectives are different, both stages rely on encoding the same user interaction sequences. 
This architectural convergence naturally motivates unifying retrieval and ranking within one Transformer backbone, which shares user sequence encoding and provides an opportunity for their supervision signals to jointly optimize the common backbone.

Recent studies~\cite{unipinrec,unisgr,unir2,onetrans_v2} have explored unifying retrieval and ranking within one Transformer backbone. Specifically,  UniPinRec~\cite{unipinrec} employs masked action modeling to support retrieval and ranking with compatible input representations, and reuses user sequence KV caches across stages. UniR$^{2}$~\cite{unir2} organizes user context, semantic ID generation, and ranking features into one decoder-only sequence, with ranking-side LoRA enabling task-specific adaptation under optimization isolation. OneTrans-V2~\cite{onetrans_v2} further extends unification to retrieval, pre-rank, and fine-rank, combining shared user context with stage-specific computation. These studies demonstrate that unified modeling could reuse user-side encoding and reduce computational costs while preserving the distinct functions of retrieval and ranking. 

Despite their encouraging process, enabling both retrieval and ranking to cooperate within a shared Transformer still faces serious \textbf{\textit{gradient conflict}}~\cite{fedosd,gradcraft,cigcn} issue, which can be attributed to the following two limitations: (1) \textbf{\textit{Retrieval and ranking have different optimization objectives, which may produce conflicting gradient directions on the shared Transformer backbone.}} 
As shown in Figure~\ref{motivation},
we perform a statistical analysis of the percentage distribution of gradient conflicts for retrieval and ranking within Transformer backbone, including Attention and FFN components.
We find that the ordinary conflict (\textit{i.e.}, cos < 0) rates in the Attention and FFN components are 45.85\% and 47.54\%, respectively. While the corresponding robust conflict (\textit{i.e.}, cos < -0.05) rates remain 39.75\% and 38.40\%, indicating substantial gradient conflict in both Attention and FFN components. 
(2) 
\textbf{\textit{When gradient conflict occurs, how can we determine a reliable direction for gradient updates?}}
In other words, a reliable gradient update direction~\cite{gupo,seed_grpo,capo} can be understood as one that reflects the overall optimization trend for both retrieval and ranking, without being affected by gradient conflict. 
This issue is particularly relevant in multi-scenario recommendation, where interaction patterns and supervision vary across scenarios. Consequently, mitigating gradient conflict while identifying reliable update directions remains a major challenge for unified retrieval and ranking.

To address the above limitations, we propose a \textit{Unified Retrieval and Ranking approach for Multi-scenario Industrial Recommendation}, termed as \textbf{LazUnirec}. 
Specifically, we construct task-shared and task-private parameters on the Transformer backbone to alleviate the gradient conflict.
Wherein, we introduce LoRA-modulated attention for the attention components, which jointly learns task-shared attention parameters and task-private low-rank adaptations for retrieval and ranking. For the FFN components, we propose task-adaptive sparse mixture-of-experts (MoE), which combines task-private with task-shared experts. 
Moreover, we present uncertainty estimation to evaluate gradient reliability of retrieval and ranking based on the historical directional stability, thereby determining a more reliable gradient update direction when gradient conflict occurs. Besides, knowledge distillation transfers candidate preference distributions from ranking to retrieval, establishing an explicit knowledge transfer path. For efficient online deployment, we further incorporate KV cache, kernel fusion, low-precision serving, and cross-scenario batching.

Our main contributions can be summarized as follows:
\vspace{-0.1cm}
\begin{itemize}
    \item We propose LazUnirec for unified retrieval and ranking in multi-scenario industrial recommendation. The LoRA-modulated attention and task-adaptive sparse MoE mitigate gradient conflict by combining task-shared and task-private parameters.
    \item We introduce uncertainty estimation to guide the model toward a more reliable direction for gradient updates when gradient conflict occurs. And knowledge distillation further establishes an explicit knowledge transfer path from ranking to retrieval.
    \item We conduct extensive offline experiments to validate the effectiveness of LazUnirec. Online A/B testing further demonstrate that LazUnirec brings considerable revenue growth (up to 4.01\% improvement in GMV) to the platform. LazUnirec has been deployed in a real-world multi-scenario recommendation system, delivering significant gains for the platform.
\end{itemize}

\section{Problem Definition}

In multi-scenario industrial recommendation, our goal is to accomplish both retrieval and ranking, as well as multi-scenario prediction, using the same Transformer backbone. Specifically, the \textbf{Retrieval} stage aims to effectively retrieve a subset of items with which the user is likely to interact. And the \textbf{Ranking} stage aims to predict multiple probabilities (\textit{i.e.}, CTR, CVR) that reflect users' preference for target (candidate) items.

Formally, let $\mathcal{U}$, $\mathcal{V}$ and $\mathcal{S}$ represent the user, item and scenario sets. The historical interaction sequence of user $u_i\in\mathcal{U}$ can be denoted as $\mathcal{H}_i=\{v_j, s_j\}_{j=1}^{|\mathcal{H}_i|}$, where $v_j\in\mathcal{V}$ is the interacted item and $s_j\in\mathcal{S}$ indicates the scenario in which the interaction occurs. Moreover, we use $\mathcal{Y}$ to denote the true labels of multi-objective prediction (\textit{i.e.}, $y^{ctr}\in\{0,1\}$, and $y^{cvr}\in\{0,1\}$) in the ranking stage. For instance, given an observed interaction pair $(u_i, v_j)$, we use $y^{ctr}_{ij}\in\{0,1\}$ to represent whether user $u_i$ clicks item $v_j$. Besides, the target (candidate) items in the ranking stage can be represented as $\mathcal{T}_i=\{v_j|v_j\in\mathcal{V}\}_{j=1}^{|\mathcal{T}_i|}$.

Thus, for a training instance of user $u_i$,
the retrieval stage adopts a sampled softmax objective as follows:
\begin{equation}
\mathcal{L}_{\mathrm{retrieval}}
=-\frac{1}{|\mathcal{P}_i|}
\sum_{j\in\mathcal{P}_i}
\log
\frac{\exp(r_{ij})}
{\exp(r_{ij})+
\sum_{n\in\mathcal{N}_{ij}}\exp(r_{in})},
\label{loss_recall}
\end{equation}
where $r_{ij}=\bm{u}_i^{\top}\bm{v}_j$ denotes the matching score between user $u_i$ and item $v_j$, $\mathcal{P}_i$ denotes the index set of positive items, and $\mathcal{N}_{ij}$ denotes the index set of sampled negative items for positive item $v_j$.

While the ranking stage minimizes the standard negative log-likelihood, which can be formulated as follows:
\begin{equation}
\begin{split}
\mathcal{L}_{\mathrm{ranking}}=-\sum^{ctr,cvr}_{xtr}[y^{xtr}_{ij}\mathrm{log}(\hat{y}^{xtr}_{ij})+(1-y^{xtr}_{ij})\mathrm{log}(1-\hat{y}^{xtr}_{ij})],
\end{split}
\label{loss_xtrs}
\end{equation}
where $y^{xtr}_{ij}$ denotes the true label of user $u_i$ to item $v_j$, and $\hat{y}^{xtr}_{ij}$ represents the predicted probability of XTRs (\textit{i.e.}, CTR, CVR).

\begin{figure*}[t!]
\begin{center}
\includegraphics[width=15.5cm]{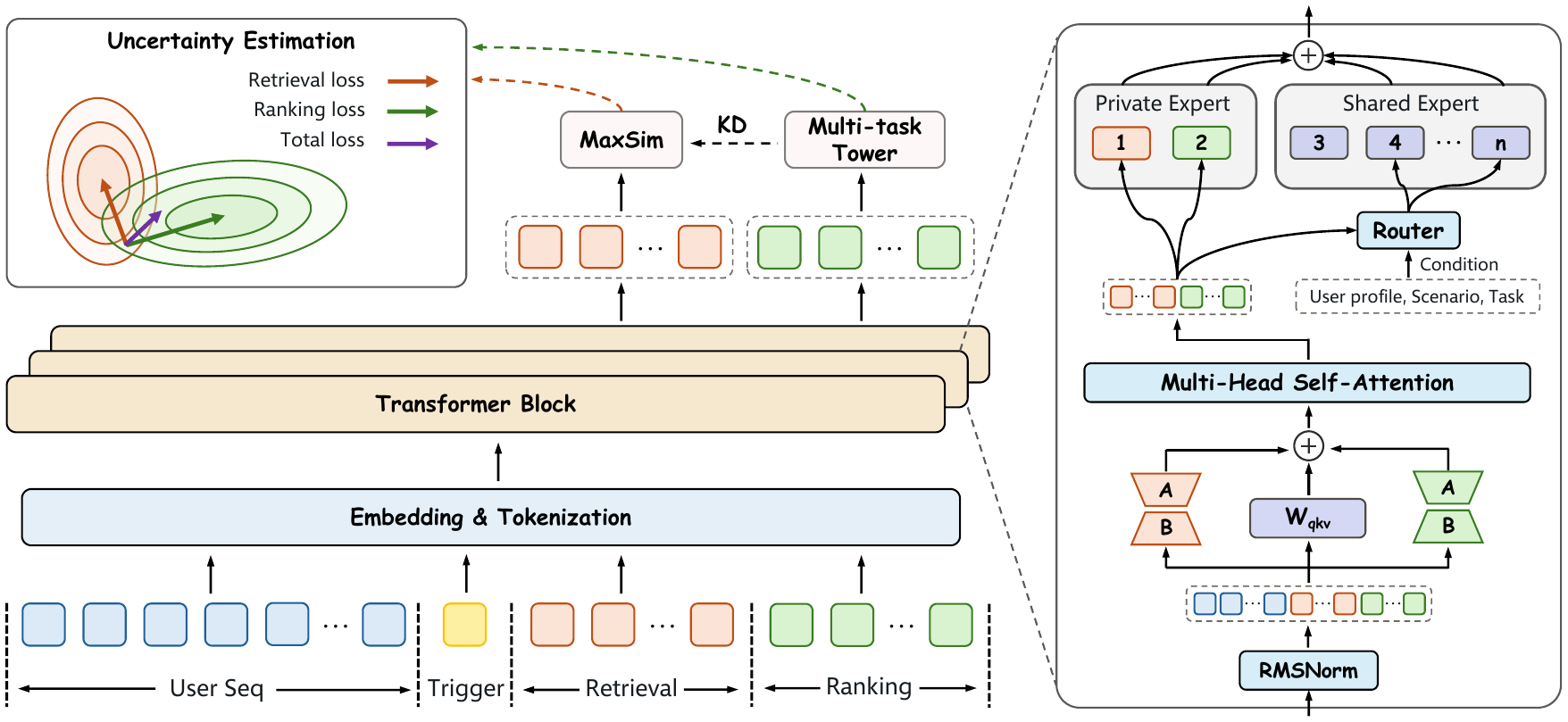}
\caption{The overall framework of LazUnirec.}
\vspace{-0.1cm}
\label{fig:main_model}
\end{center}
\end{figure*}

\section{Methodology}

This section introduced our proposed LazUnirec, which implements a unified retrieval and ranking approach for multi-scenario industrial recommendation, as depicted in Figure~\ref{fig:main_model}. Specifically, LazUnirec mainly consists of five parts: 1) the \textbf{unified sequence tokenization} aims to integrate user interaction sequence tokens, retrieval and ranking tokens from each scenario into a unified sequence, 2) the \textbf{LoRA-modulated attention} draws inspiration from LoRA~\cite{lora} by constructing task-private parameters for retrieval and ranking to maintain task specificity while sharing the underlying attention parameters, 3) the \textbf{task-adaptive sparse MoE} further mitigates the gradient conflict by constructing task-shared and task-private experts, 4) the \textbf{uncertainty estimation} aims to determine a reliable direction for gradient updates when gradient conflict occurs, and 5) the \textbf{knowledge distillation} establishes an explicit knowledge transfer path from ranking to retrieval. Details of each component are demonstrated in the following parts.

\subsection{Unified Sequence Tokenization}

To unify retrieval and ranking across scenarios, we organize the input of user $u_i$ into a unified sequence $\bm{X}_i
=[\bm{H}_i\Vert\bm{g}_i\Vert\bm{R}_i\Vert\bm{T}_i]$, which consists of interaction sequence tokens $\bm{H}_i$, a trigger token $\bm{g}_i$, retrieval tokens $\bm{R}_i$, and ranking tokens $\bm{T}_i$.

\subsubsection{\textbf{User Sequence Tokens}}
\label{user_sequence_tokens}
Given interacted sequence item $(v_j,s_j)\in\mathcal{H}_i$, we construct sequence token $\bm{h}_j$ by combining id-based (\textit{i.e.}, item ID), side info-based (\textit{i.e.}, item’s attribute information), and scenario features as follows:
\begin{equation}
\begin{split}
\bm{h}_j
&=[\bm{e}_j^{id}\Vert\bm{e}_j^{side}\Vert\bm{e}_j^{scenario}],\\
\bm{H}_i
&=[\bm{h}_1,\bm{h}_2,\ldots,\bm{h}_{|\mathcal{H}_i|}],
\end{split}
\label{eq:history_token}
\end{equation}
where $\Vert$ denotes the concatenation operation.


\subsubsection{\textbf{Trigger Token}}
In our multi-scenario industrial recommendation,
scenarios are categorized as trigger-based when a trigger item is available and trigger-free otherwise. Thus, we append a trigger token $\bm{g}_i$ to distinguish these two types of scenarios. Specifically, trigger-based scenarios encode the trigger item to obtain $\bm{g}_i$ similar to Eq.~(\ref{eq:history_token}). For trigger-free scenarios, a learnable placeholder replaces the trigger item. Accordingly, the trigger token summarizes the user sequence through attention and provides scenario context for constructing subsequent retrieval tokens.

\subsubsection{\textbf{Retrieval Tokens}}
We introduce multiple learnable retrieval queries $\{\bm{b}_k\}_{k=1}^{K}$ to capture diverse user preferences. Each query is combined with the trigger representation to construct a retrieval token $\bm{r}_{k}$, which can be formulated as follows:
\begin{equation}
\begin{split}
\bm{r}_{k}
&=
\bm{g}_i+\bm{b}_k,\\
\bm{R}_i
&=[\bm{r}_1,\bm{r}_2,\ldots,\bm{r}_{K}],
\end{split}
\label{eq:retrieval_tokens}
\end{equation}
These retrieval tokens form $\bm{R}_i$ and extract multiple user preference representations in parallel~\cite{mima}.

\subsubsection{\textbf{Ranking Tokens}}
Industrial logs typically contain multiple target items within the same request. 
Traditional industrial recommendation systems construct a separate sample for each target item, leading to the redundant encoding of the same user sequence.
Thus, we organize all target items within the same request~\cite{lazformer} to share the storage, transfer, and encoding of user sequence. Specifically, for each target item $v_j$, we construct ranking token $\bm{t}_j$ similar to Eq.~(\ref{eq:history_token}) as follows:
\begin{equation}
\begin{split}
\bm{t}_j
&=[\bm{e}_j^{id}\Vert\bm{e}_j^{side}\Vert\bm{e}_j^{scenario}],\\
\bm{T}_i
&=[\bm{t}_1,\bm{t}_2,\ldots,\bm{t}_{|\mathcal{T}_i|}],
\end{split}
\label{eq:ranking_tokens}
\end{equation}

Consequently, the unified input sequence for LazUnirec can be formulated as follows:
\begin{equation}
\begin{split}
\bm{X}_i
&=[\bm{H}_i\Vert\bm{g}_i\Vert\bm{R}_i\Vert\bm{T}_i],\\
&=[
\bm{h}_1,\ldots,\bm{h}_{|\mathcal{H}_i|}~\Vert~\bm{g}_i~\Vert~\bm{r}_1,\ldots,\bm{r}_{K}~\Vert~ \bm{t}_{1},\ldots,\bm{t}_{|\mathcal{T}_i|}
],
\label{eq:ranking_input_sequence}
\end{split}
\end{equation}

\subsection{Transformer Backbone}

Existing studies~\cite{unipinrec,unisgr,unir2,onetrans_v2} unify retrieval and ranking through a shared Transformer backbone, while their different optimization objectives may produce conflicting gradient on shared parameters. Thus, we introduce task-shared and task-private parameters into both the attention and feed-forward networks (FFN) components. 
These private parameters provide additional capacity to accommodate task-specific patterns, alleviating gradient conflict while preserving knowledge sharing between retrieval and ranking.

\begin{figure}[t!]
\begin{center}
\includegraphics[width=4.5cm]{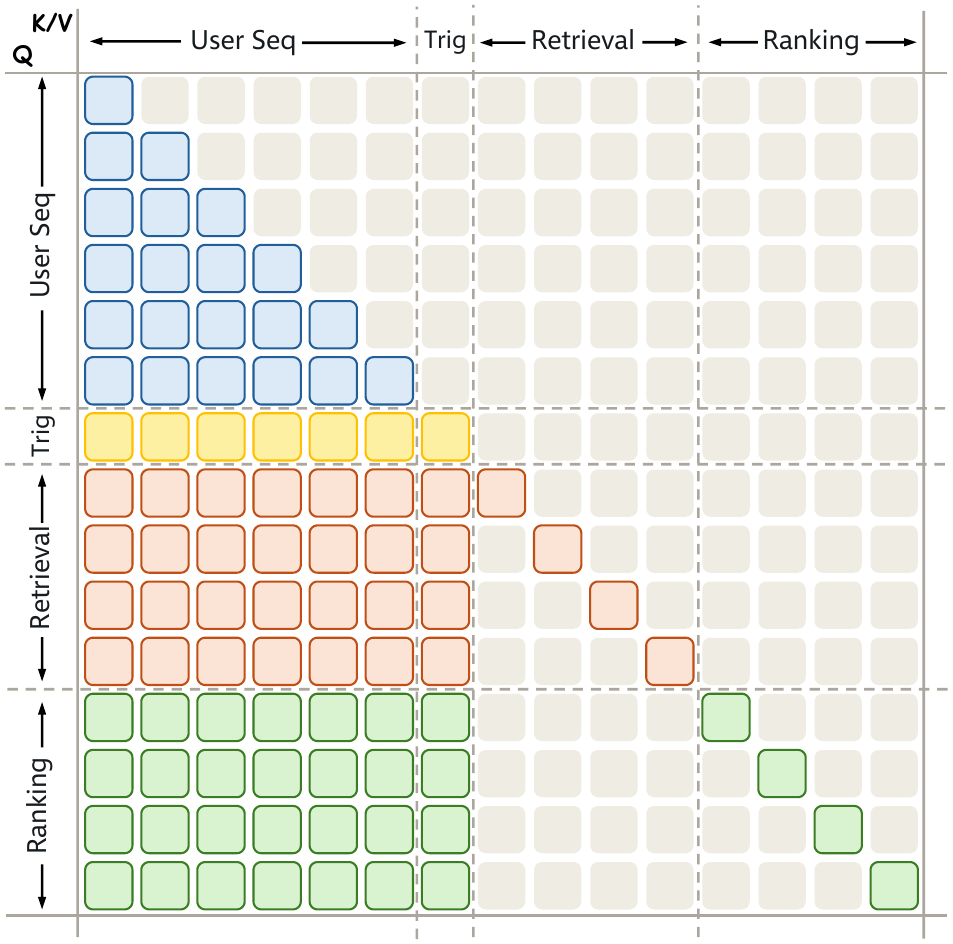}
\caption{An illustration of the attention mask.}
\vspace{-0.4cm}
\label{fig:attention_mask}
\end{center}
\end{figure}

\subsubsection{\textbf{LoRA-modulated Attention}}

We introduce task-private low-rank adaptations~\cite{lora} to the query, key, and value projections. Specifically, for retrieval tokens $\bm{R}_k$ and ranking tokens $\bm{T}_j$, we augment the shared query, key, and value projections with task-private low-rank residual as follows:
\begin{equation}
\begin{split}
\bm{p}_{k}^{\mathrm{retrieval}}
&=(\bm{W}_{p}+\Delta\bm{W}_{p}^{\mathrm{ret}})\bm{R}_{k},\\
\bm{p}_{j}^{\mathrm{ranking}}
&=(\bm{W}_{p}+\Delta\bm{W}_{p}^{\mathrm{rank}})\bm{T}_{j},\\
\Delta\bm{W}_{p}^{\tau}
=\bm{B}_{p}^{\tau}\bm{A}_{p}^{\tau},&\quad\tau\in\{\mathrm{retrieval},\mathrm{ranking}\},
\end{split}
\label{eq:lora_projection}
\end{equation}
where $p\in\{Q,K,V\}$ denotes the projection type, and $\bm{A}_{p}^{\tau}$ and $\bm{B}_{p}^{\tau}$ are task-private low-rank matrices. User sequence tokens $\bm{H}_i$ and the trigger token $\bm{g}_i$ use only the shared projections. Overall, both task-shared and task-private parameters are jointly optimized, allowing the two tasks to share knowledge while accommodating their distinct objectives.

Moreover, we introduce a structured attention mask $\bm{M}_i$ to control information flow within the unified sequence $\bm{X}_i$, as shown in Figure~\ref{fig:attention_mask}. Specifically, user sequence tokens attend to themselves and preceding interactions, while the trigger token attends to the user sequence and itself. Each retrieval or ranking token attends only to the user sequence, the trigger token, and itself, preserving the isolation of different retrieval queries and ranking candidates.

The attention operation for each head is formulated as follows:
\begin{equation}
\bm{O}_i
=
\operatorname{Softmax}\left(
\frac{\bm{Q}_i\bm{K}_i^{\top}}{\sqrt{d_h}}
+\bm{M}_i
\right)\bm{V}_i,
\label{eq:masked_attention}
\end{equation}
where $\bm{Q}_i$, $\bm{K}_i$, and $\bm{V}_i$ denote the query, key, and value matrices, and $d_h$ is the head dimension. Through LoRA-modulated attention, LazUnirec jointly learns task-shared attention parameters and task-private low-rank adaptations for retrieval and ranking, preserving cross-task knowledge sharing while accommodating task-specific objectives to mitigate gradient conflict.

\subsubsection{\textbf{Task-adaptive Sparse MoE}}

To further alleviate gradient conflict in the FFN components, we introduce a task-adaptive sparse MoE that combines task-shared and task-private experts. Specifically, the MoE consists of a shared expert pool and one task-private expert for each task, enabling cross-task knowledge sharing while preserving task specificity.

Given the retrieval or ranking token $\bm{x}$ produced by the LoRA-modulated attention,
its task-private expert is always activated. Meanwhile, a learnable router selects shared experts based on the condition information, including token representation, user profile, scenario, and task, which can be formulated as follows:
\begin{equation}
\begin{split}
&\bm{a}^{\tau}
=
\operatorname{Router}(
\bm{x},
\bm{e}_i^{profile},
\bm{e}_i^{scenario},
\bm{e}_{\tau}^{task}
),
\end{split}
\label{eq:router_scores}
\end{equation}
where $\bm{a}^{\tau}$ denotes the routing scores over shared experts for the current token of task $\tau\in\{\mathrm{retrieval},\mathrm{ranking}\}$.

The router then selects the top-$k$ shared experts as follows:
\begin{equation}
\begin{split}
\mathcal{D}^{\tau}&=\operatorname{TopK}(\bm{a}^{\tau},k),\\
\quad
w_e^\tau&=\frac{\exp(a_e^\tau)}
{\sum_{m\in\mathcal{D}^{\tau}}\exp(a_m^\tau)},
\quad e\in\mathcal{D}^{\tau},
\end{split}
\label{eq:expert_routing}
\end{equation}
where $\mathcal{D}^{\tau}$ contains the selected expert indices, and $w_e^\tau$ denotes the normalized routing weight assigned to shared expert $e$ for the current token of task $\tau$. We set $k=2$ in our implementation.

Ultimately, The MoE combines the outputs of task-private expert with the weighted outputs of the selected shared experts:
\begin{equation}
\operatorname{MoE}^{\tau}(\bm{x})
=
\operatorname{E}_{\mathrm{private}}^{\tau}(\bm{x})
+
\sum_{e\in\mathcal{D}^{\tau}}
w_e^\tau\operatorname{E}_{e}^{\mathrm{shared}}(\bm{x}),
\label{eq:task_adaptive_moe}
\end{equation}

Consequently, task-private experts provide dedicated capacity for task-specific patterns, while selectively activated shared experts facilitate knowledge sharing between retrieval and ranking. The router adapts expert selection to user profiles, scenarios, and tasks, while sparse activation controls computational cost.

\subsection{Uncertainty Estimation}

In addition to alleviating gradient conflict by structuring task-shared and task-private parameters, we further propose an uncertainty estimation to evaluate the gradient reliability~\cite{gupo,seed_grpo,capo} of retrieval and ranking, thereby guiding the model toward a more reliable gradient update direction when gradient conflict occurs. Specifically, we estimate gradient uncertainty based on historical directional stability, assigning relatively higher weights to stable gradients and lower weights to fluctuating ones.

For scenario $s\in\mathcal{S}$ and task $\tau\in\{\mathrm{retrieval},\mathrm{ranking}\}$, let $\bm{\gamma}_s^\tau$ denote a normalized gradient representation obtained by periodically probing a small subset of shared parameters, where the normalization focuses the estimation on gradient directions rather than their magnitudes. We maintain exponential moving averages (EMA) to estimate the variance and convert it into gradient evidence:
\begin{equation}
\begin{split}
(\bm{\sigma}_{s}^{\tau})^2
&=[
\operatorname{EMA}((\bm{\gamma}_{s}^{\tau})^2)
-(\operatorname{EMA}(\bm{\gamma}_{s}^{\tau}))^2
]_{+},\\
\bm{e}_{s}^{\tau}
&=\min(
e_{\max},
((\bm{\sigma}_{s}^{\tau})^2+\epsilon)^{-\frac{1}{2}}
),
\end{split}
\label{eq:gradient_evidence}
\end{equation}
where $[x]_{+}=\max(x,0)$, $\epsilon$ and $e_{\max}$ ensure numerical stability. Smaller historical variance indicates more stable gradient directions and yields stronger evidence.

We then aggregate the evidence to estimate uncertainty and determine the corresponding loss weight as follows:
\begin{equation}
\begin{split}
\mathcal{E}_s^\tau
&=\frac{1}{1+\operatorname{Mean}(\bm{e}_s^\tau)},\\
\omega_s^\tau
&=\frac{1-\mathcal{E}_s^\tau}{\hat{(1-\mathcal{E}_s)}_{\operatorname{batch}}}
,
\end{split}
\label{eq:uncertainty_weight}
\end{equation}
where $\operatorname{Mean}$ averages the evidence entries, $1-\mathcal{E}_s^\tau$ represents gradient reliability, and $\hat{(1-\mathcal{E}_s)}_{\operatorname{batch}}$ is the mean reliability over retrieval and ranking observations in the current mini-batch.

Ultimately, we apply the reliability weights to the task losses to obtain the uncertainty weighted objective function:
\begin{equation}
\mathcal{L}_{\mathrm{ue}}
=
\sum_{\tau}
\operatorname{sg}(\omega_s^{\tau})
\mathcal{L}^{\tau},
\label{eq:uncertainty_objective}
\end{equation}
where $\mathcal{L}^{\tau}$ denotes the retrieval or ranking loss, and $\operatorname{sg}$ denotes stop-gradient, ensuring that the reliability weights do not participate in gradient computation.

Through the uncertainty estimation, gradients with more stable historical directions receive greater relative influence on shared parameters update, guiding joint optimization toward more reliable directions when gradient conflict occurs.

\subsection{Knowledge Distillation}

To further facilitate cross-task knowledge transfer, we introduce listwise knowledge distillation from ranking to retrieval. Specifically, the ranking branch captures fine-grained user-item interactions and provides relative preference signals over target items within the same request. We transfer these signals to retrieval by matching their candidate preference distributions.

For target items $\mathcal{T}_i$, we construct request-level candidate lists by grouping candidate items with the same timestamp. The retrieval score for distillation is calculated as follows:
\begin{equation}
\tilde{r}_{ij}
=
\max_{1\leq k\leq K}
\left(
\pi_{ik}\bm{u}_{ik}^{\top}\bm{v}_j
\right),
\label{eq:distillation_retrieval_score}
\end{equation}
where $\bm{u}_{ik}$ is the user interest representation (\textit{i.e.}, the Transformer output of the $k$-th retrieval token $\bm{r}_k$), and $\pi_{ik}$ denotes the corresponding interest probability.

Let $\bm{z}_i^{ctr}$ collect the ranking CTR logits over the candidate list, and $\tilde{\bm{r}}_i$ collect the corresponding retrieval scores, we construct teacher and student distributions as follows:
\begin{equation}
\begin{split}
\bm{p}_i^{\mathrm{ranking}}
&=
\operatorname{Softmax}(
\frac{\operatorname{sg}(\bm{z}_i^{ctr})}
{T_{\mathrm{ranking}}}
),\\
\bm{p}_i^{\mathrm{retrieval}}
&=
\operatorname{Softmax}(
\frac{\tilde{\bm{r}}_i}
{T_{\mathrm{retrieval}}}
),
\end{split}
\label{eq:distillation_distributions}
\end{equation}
where $T_{\mathrm{ranking}}$, $T_{\mathrm{retrieval}}$ are temperature hyperparameters controlling distribution sharpness, and stop-gradient blocks backpropagation through the ranking teacher.

We then minimize the KL divergence from the ranking distribution to the retrieval distribution as follows:
\begin{equation}
\begin{split}
\mathcal{L}_{\mathrm{kd}}
=
\sum_{v_j\in\mathcal{T}_i}
p_{ij}^{\mathrm{ranking}}
\log
\frac{p_{ij}^{\mathrm{ranking}}}
{p_{ij}^{\mathrm{retrieval}}},
\end{split}
\label{eq:distillation_objective}
\end{equation}

Knowledge distillation updates the retrieval representations and interest probabilities through the student branch, enabling explicit knowledge transfer from ranking to retrieval.

Overall, LazUnirec is optimized by combining the uncertainty weighted loss, knowledge distillation, and MoE regularization, which can be formulated as follows:
\begin{equation}
\begin{split}
\mathcal{L}
&=\mathcal{L}_{\mathrm{ue}}
+\lambda_{\mathrm{kd}}\mathcal{L}_{\mathrm{kd}}
+\lambda_z\mathcal{L}_z\\
&=\operatorname{sg}(\omega_s^{\mathrm{retrieval}})
\mathcal{L}_{\mathrm{retrieval}}
+\operatorname{sg}(\omega_s^{\mathrm{ranking}})
\mathcal{L}_{\mathrm{ranking}}\\
&\quad+\lambda_{\mathrm{kd}}\mathcal{L}_{\mathrm{kd}}
+\lambda_z\mathcal{L}_z,
\end{split}
\label{eq:overall_objective}
\end{equation}
where $\mathcal{L}_z$ denotes the MoE router z-loss, $\lambda_{\mathrm{kd}}$ and $\lambda_z$ control the contributions of distillation and MoE regularization, respectively.

\subsection{Online Deployment}

In online inference, We deploy two separate models to serve retrieval and ranking.
Specifically, we optimize LazUnirec through KV cache, kernel fusion, low-precision serving, and cross-scenario batching. These optimizations reduce repeated sequence encoding, lower computational and memory overhead, and improve GPU utilization by jointly processing requests from different scenarios.

\subsubsection{\textbf{KV Cache}}

We construct a user KV cache during the retrieval stage, storing the layer-wise keys and values of user sequences. Within the same request, the ranking stage reuses this cache while performing its task-specific computation. The cache is further reused across requests when user sequence and scenario conditions remain unchanged. The KV cache enables cross-task and cross-request reuse, reducing redundant user sequence encoding.

\subsubsection{\textbf{Kernel Fusion}}

In order to reduce memory traffic and kernel launch overhead, we combine a series of operations that share the same inputs. Specifically, the shared query, key, and value projection weights are concatenated so that all three projections are computed through one matrix multiplication. Within SwiGLU (\textit{i.e.}, expert), we fuse the gate/up projections, SiLU activation, and element-wise multiplication to reduce intermediate tensor materialization.
For sparse MoE, tokens assigned to the same expert are organized contiguously in GPU memory. We apply Grouped GEMM (\textit{i.e.}, general matrix multiplication) to both the gate/up and down projections, executing the corresponding matrix multiplications across multiple experts in a single kernel launch per group. This organization reduces separate expert launches and improves the utilization of GPU computation.

\subsubsection{\textbf{Low-Precision Serving}}

We adopt FP16 as the default serving precision throughout LazUnirec and selectively use FP8 for the matrix multiplications within SwiGLU, including the gate/up and down projections. 
This configuration targets the computational and memory costs of FFN execution while restricting FP8 quantization to a specific part of the model.

\subsubsection{\textbf{Cross-scenario Batching}}

LazUnirec serves multiple scenarios simultaneously in online inference. Thus, we aggregate online requests across scenarios to improve GPU utilization. Requests arriving within a very short time window are grouped into the same inference batch, regardless of their originating scenarios. This strategy combines traffic from different scenarios to form larger batches, thereby minimizing additional waiting time.

\section{Experiments}
In this section, we conduct extensive offline experiments and online A/B testing on our multi-scenario industrial recommendation system to answer the following research questions:
\begin{itemize}
    \item \textbf{RQ1:} How does our LazUnirec perform compared to existing retrieval and ranking baselines?
    \item \textbf{RQ2:} How does different components of our LazUnirec contribute to the overall model performance?
    \item \textbf{RQ3:} Do the LoRA-modulated attention and task-adaptive sparse MoE proposed in LazUnirec really alleviate gradient conflict?
    \item \textbf{RQ4:} Does uncertainty estimation proposed in LazUnirec really guide the model toward a more reliable gradient update direction when gradient conflict occurs?
    \item \textbf{RQ5:} Does training data from muti-scenario really improve experimental performance?
    \item \textbf{RQ6:} How is the online efficiency of our LazUnirec?
    \item \textbf{RQ7:} How does LazUnirec perform in real-world online A/B testing under multi-scenario industrial deployment settings?
\end{itemize}

\begin{table*}[t]
\centering
\caption{Offline ranking performance in terms of AUC and GAUC, where the best results are in bold and the best baseline ones are underlined. \% Improve represents the relative improvements of LazUnirec over the best baseline.}
\label{exp_ranking_offline}
\setlength{\tabcolsep}{2.0pt}
\begin{tabular*}{\textwidth}
{@{\extracolsep{\fill}}@{}lcccccccccccc@{}}
\toprule
&
\multicolumn{4}{c}{\textbf{Scenario 1}} &
\multicolumn{4}{c}{\textbf{Scenario 2}} &
\multicolumn{4}{c}{\textbf{Scenario 3}} \\
\cmidrule(r){2-5}\cmidrule(r){6-9}\cmidrule(r){10-13}
\textbf{Models} &
\multicolumn{2}{c}{\textbf{CTR}} & \multicolumn{2}{c}{\textbf{CVR}} &
\multicolumn{2}{c}{\textbf{CTR}} & \multicolumn{2}{c}{\textbf{CVR}} &
\multicolumn{2}{c}{\textbf{CTR}} & \multicolumn{2}{c}{\textbf{CVR}} \\
\cmidrule(r){2-3}\cmidrule(r){4-5}
\cmidrule(r){6-7}\cmidrule(r){8-9}
\cmidrule(r){10-11}\cmidrule(r){12-13}
&
\textbf{AUC} & \textbf{GAUC} & \textbf{AUC} & \textbf{GAUC} &
\textbf{AUC} & \textbf{GAUC} & \textbf{AUC} & \textbf{GAUC} &
\textbf{AUC} & \textbf{GAUC} & \textbf{AUC} & \textbf{GAUC} \\
\midrule
Transformer
& 0.7623 & 0.6678 & 0.8923 & 0.6904
& 0.7642 & 0.6765 & 0.8454 & 0.5750
& 0.7239 & 0.6486 & 0.8506 & 0.5701 \\
UniPinRec
& 0.7630 & 0.6677 & 0.8929 & 0.6907
& 0.7650 & 0.6777 & 0.8457 & 0.5761
& 0.7245 & 0.6495 & 0.8506 & 0.5689 \\
$\operatorname{UniR^2}$
& 0.7596 & 0.6635 & 0.8900 & 0.6913
& 0.7606 & 0.6717 & 0.8424 & 0.5751
& 0.7198 & 0.6444 & 0.8471 & 0.5691 \\
UniSGR
& 0.7630 & 0.6687 & \underline{0.8936} & \underline{0.6934}
& 0.7650 & 0.6775 & 0.8471 & 0.5730
& 0.7249 & 0.6496 & 0.8517 & \underline{0.5707} \\
OneTrans-V2
& \underline{0.7634} & \underline{0.6695} & 0.8935 & 0.6928
& \underline{0.7654} & \underline{0.6781}
& \underline{0.8478} & \underline{0.5778}
& \underline{0.7257} & \underline{0.6499} & \underline{0.8524} & 0.5705 \\
\midrule
LazUnirec
& \textbf{0.7655} & \textbf{0.6722} & \textbf{0.8966} & \textbf{0.6972}
& \textbf{0.7678} & \textbf{0.6828} & \textbf{0.8502} & \textbf{0.5803}
& \textbf{0.7274} & \textbf{0.6518} & \textbf{0.8555} & \textbf{0.5733} \\
\% \textit{Improve} $\uparrow$
& +0.28\% & +0.40\% & +0.34\% & +0.55\%
& +0.31\% & +0.69\% & +0.28\% & +0.43\%
& +0.23\% & +0.29\% & +0.36\% & +0.46\% \\
\bottomrule
\end{tabular*}
\end{table*}

\begin{table*}[t]
\centering
\caption{Offline retrieval performance in terms of HR@100, HR@500, and HR@1000, where the best results are in bold and the best baseline ones are underlined. \% Improve represents the relative improvements of LazUnirec over the best baseline.}
\label{exp_retrieval_offline}
\setlength{\tabcolsep}{2.0pt}
\begin{tabular*}{\textwidth}
{@{\extracolsep{\fill}}@{}lccccccccc@{}}
\toprule
\multirow{2.5}{*}{\textbf{Models}} &
\multicolumn{3}{c}{\textbf{Scenario 1}} &
\multicolumn{3}{c}{\textbf{Scenario 2}} &
\multicolumn{3}{c}{\textbf{Scenario 3}} \\
\cmidrule(r){2-4}\cmidrule(r){5-7}\cmidrule(r){8-10}
&
\textbf{HR@100} & \textbf{HR@500} & \textbf{HR@1000} &
\textbf{HR@100} & \textbf{HR@500} & \textbf{HR@1000} &
\textbf{HR@100} & \textbf{HR@500} & \textbf{HR@1000} \\
\midrule
Transformer
& 0.3275 & 0.5502 & 0.6402
& 0.5508 & 0.7556 & 0.8156
& 0.7954 & 0.9009 & 0.9207 \\
UniPinRec
& 0.3264 & 0.5503 & 0.6416
& 0.5532 & 0.7583 & 0.8189
& 0.7980 & 0.9034 & 0.9236 \\
$\operatorname{UniR^2}$
& 0.3305 & \underline{0.5553} & 0.6447
& 0.5536 & \underline{0.7596} & 0.8198
& 0.7970 & 0.9047 & 0.9214 \\
UniSGR
& 0.3305 & 0.5508 & 0.6433
& 0.5541 & 0.7567 & 0.8171
& 0.7967 & \underline{0.9059} & \underline{0.9239} \\
OneTrans-V2
& \underline{0.3346} & 0.5544 & \underline{0.6484}
& \underline{0.5557} & 0.7591 & \underline{0.8203}
& \underline{0.7984} & 0.9050 & 0.9221 \\
\midrule
LazUnirec
& \textbf{0.3620} & \textbf{0.5892} & \textbf{0.6802}
& \textbf{0.5906} & \textbf{0.7873} & \textbf{0.8432}
& \textbf{0.8181} & \textbf{0.9169} & \textbf{0.9389} \\
\% \textit{Improve} $\uparrow$
& +8.19\% & +6.10\% & +4.90\%
& +6.28\% & +3.65\% & +2.79\%
& +2.47\% & +1.21\% & +1.62\% \\
\bottomrule
\end{tabular*}
\end{table*}

\subsection{Experimental Settings}
\subsubsection{\textbf{Datasets}}

We evaluate the offline performance of LazUnirec using large-scale industrial data from a real-world e-commerce platform, including multiple trigger-free (\textit{i.e.}, scenario 1) and trigger-based (\textit{i.e.}, scenario 2, 3) recommendation scenarios. 
For retrieval and ranking, we organize training samples at the request level, where each sample contains the user's chronologically ordered interaction sequences, the current scenario, and candidate items with their observed feedback.
We use logs from 25 consecutive days for training and the subsequent day for testing.

\subsubsection{\textbf{Baselines}}

We compare \textbf{LazUnirec} against a vanilla \textbf{Transformer} and recent representative approaches that unify retrieval and ranking, including \textbf{UniPinRec}~\cite{unipinrec}, \textbf{UniR$\bm{^2}$}~\cite{unir2}, \textbf{UniSGR}~\cite{unisgr}, and \textbf{OneTrans-V2}~\cite{onetrans_v2}.

\subsubsection{\textbf{Evaluation Metrics}}

Follwing previous methods~\cite{unipinrec,unir2,onetrans_v2},
we evaluate the effectiveness of LazUnirec and baselines on both retrieval and ranking. Specifically, we report HR@$N$ (Hit Rate) with $N\in\{100,500,1000\}$ for retrieval, and AUC, GAUC for ranking.

\subsubsection{\textbf{Implementation Details}}

We implement LazUnirec with PyTorch and optimize using AdamW~\cite{adam} on NVIDIA H100 GPUs. All baselines and our LazUnirec share the same backbone configuration, which consists of 5 Transformer layers with hidden size 256, and 4 attention heads. The maximum history length is 2,048, and all models are trained for one epoch.

Each request contains up to 15 ranking tokens and $K=8$ retrieval queries. The LoRA ranks for retrieval and ranking are set to 8 and 16, respectively. Each task-adaptive sparse MoE contains 14 shared experts and 1 private expert per task, activating the task-private expert and the top-$k$ shared experts with $k=2$. The intermediate dimension of each expert is 1,024, and the MoE regularization coefficient is set to $\lambda_z=0.001$.

For uncertainty estimation, we probe gradients every 100 optimization steps, and use an EMA decay of 0.99. The reliability weights $\omega_s^\tau$ are clipped to $[0.8,1.2]$.
For knowledge distillation, the temperatures $T_{\mathrm{ranking}}$ and $T_{\mathrm{retrieval}}$ are set to 0.5 and 0.01, respectively, with $\lambda_{\mathrm{kd}}$ increases linearly to 0.05 at microstep 5,000.

\subsection{Offline Performance (RQ1)}

In this section, we compare the offline performance of LazUnirec against baselines in Table~\ref{exp_ranking_offline} and~\ref{exp_retrieval_offline}, from which we can summarize the following observations.

LazUnirec consistently outperforms all baselines in both retrieval and ranking across the three scenarios. 
This can be attributed to LoRA-modulated attention and task-adaptive sparse MoE, which preserve task-specific capacity while maintaining cross-task knowledge sharing. Moreover, uncertainty estimation assigns relatively higher weights to gradients with more stable historical direction, guiding the model toward a more reliable gradient update direction when gradient conflict occurs.
Another observation is that, $\operatorname{UniR^2}$ performs worst on most ranking metrics, even falling below the vanilla Transformer across all scenarios. The explanation is that its stop-gradient mechanism blocks ranking gradients from updating the shared base Transformer parameters. Although task-specific components retain ranking adaptability, this operation limits the adaptation of shared parameters to ranking objectives. In contrast, LazUnirec allows both retrieval and ranking to optimize shared parameters while alleviating the gradient conflict.
Besides, scenario-aware expert routing and uncertainty estimation accommodate differences in scenarios while benefiting from shared knowledge, and the multi-scenario modeling enables LazUnirec to learn transferable user preferences from interactions across scenarios. We further analyze the multi-scenario modeling in Section~\ref{multi_scenario_analysis}.

\begin{table}[t]
\centering
\caption{Ablation study on Scenario 1.}
\label{exp_ablation}
\setlength\tabcolsep{2.0pt}
\begin{tabular*}{0.49 \textwidth}
{@{\extracolsep{\fill}}@{}lccccccc@{}}
\toprule
\multirow{2.5}{*}{\bf{Models}} &
\multicolumn{2}{c}{\bf{CTR}} &
\multicolumn{2}{c}{\bf{CVR}} &
\multicolumn{3}{c}{\bf{HR}} \\
\cmidrule(r){2-3}\cmidrule(r){4-5}\cmidrule(r){6-8}
&
\bf{AUC} & \bf{GAUC} &
\bf{AUC} & \bf{GAUC} & \bf{@100} & \bf{@500} & \bf{@1000} \\
\midrule
\textit{w/o} LoRA & 0.7646 & 0.6709 & 0.8955 & 0.6952 & 0.3532 & 0.5819 & 0.6732 \\
\textit{w/o} SMoE & 0.7634 & 0.6692 & 0.8943 & 0.6923 & 0.3408 & 0.5659 & 0.6531 \\
\textit{w/o} UE & 0.7643 & 0.6704 & 0.8951 & 0.6941 & 0.3528 & 0.5755 & 0.6682 \\
\textit{w/o} KD & 0.7653 & 0.6719 & 0.8969 & 0.6977 & 0.3502 & 0.5781 & 0.6666 \\
LazUnirec & \textbf{0.7655} & \textbf{0.6722} & \textbf{0.8966} & \textbf{0.6972} & \textbf{0.3620} & \textbf{0.5892} & \textbf{0.6802} \\
\bottomrule
\end{tabular*}
\end{table}

\begin{figure}[t]
\setlength{\abovecaptionskip}{0.cm}
	\begin{center}
        \subfigure
        {\begin{minipage}[b]{.47\linewidth}
        \centering
        \includegraphics[scale=0.205]{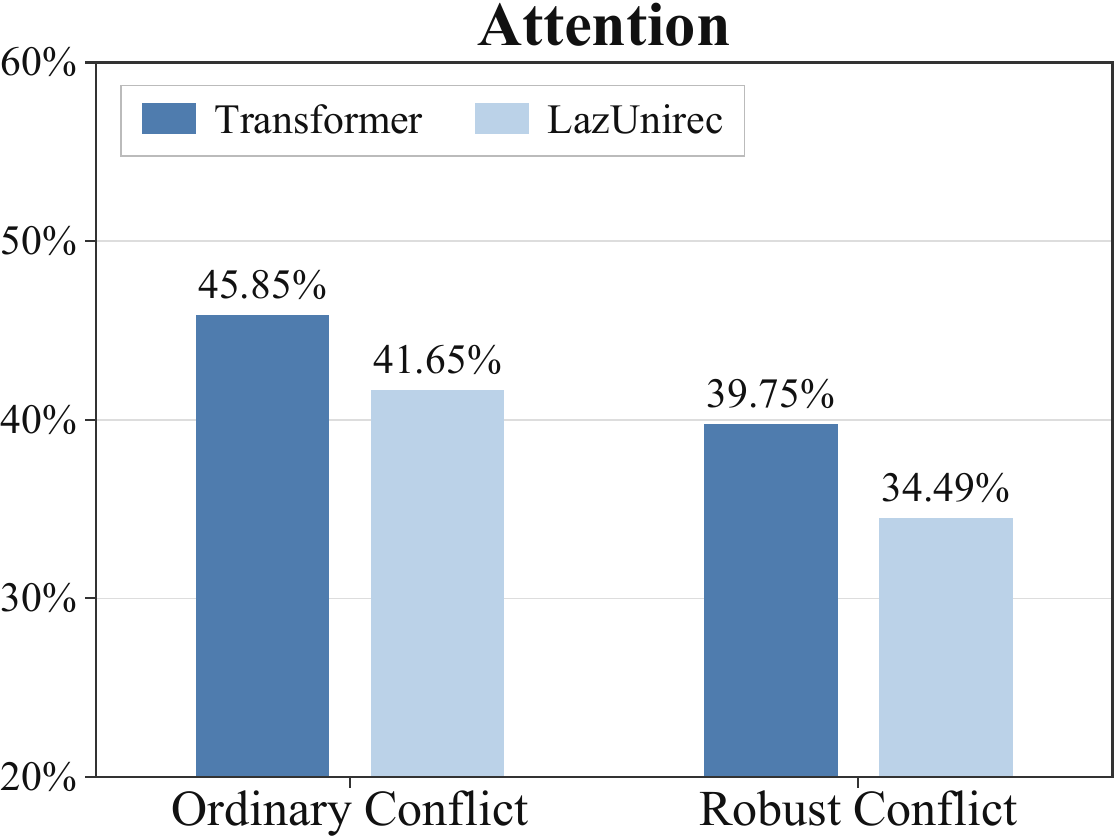}
        \end{minipage}}
        \subfigure
        {\begin{minipage}[b]{.47\linewidth}
        \centering
        \includegraphics[scale=0.205]{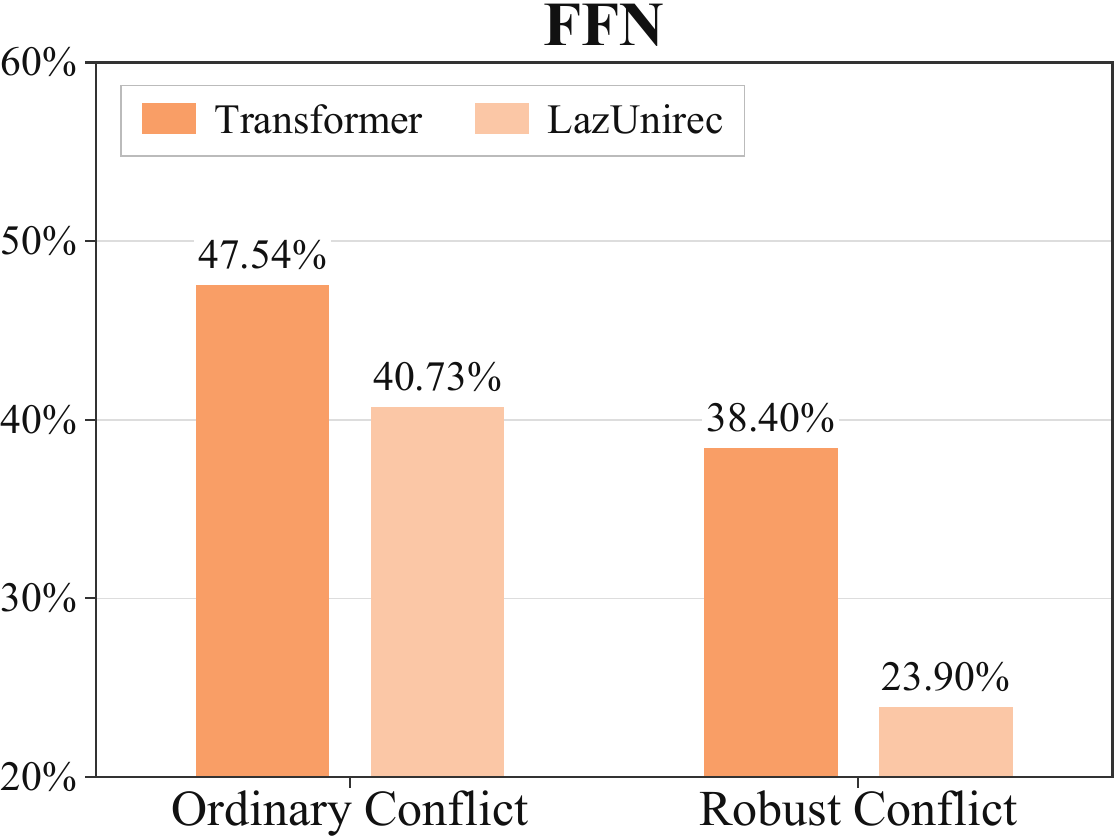}
        \end{minipage}}
        
	\caption{Statistics on the percentage of gradient conflicts within Attention and FFN components for  vanilla Transformer and LazUnirec, respectively.}
	\label{gradient_conflict_analysis}
	\end{center}
\vspace{-0.1cm}
\end{figure}

\subsection{Ablation Study (RQ2)}

In order to evaluate the effectiveness of each component in LazUnirec, we construct the ablation study on scenario 1. Specifically, we develop four variants to establish a comparative analysis, which are outlined as follows:
\begin{itemize}
    \item \textit{w/o} LoRA: This variant excludes the LoRA-modulated attention.
    \item \textit{w/o} SMoE: This variant excludes the task-adaptive sparse MoE.
    \item \textit{w/o} UE: This variant removes the uncertainty estimation.
    \item \textit{w/o} KD: In this variant, we remove the knowledge distillation.
\end{itemize}
Form the results shown in Table~\ref{exp_ablation},
we find that the removal of either LoRA-modulated attention or task-adaptive sparse MoE degrades both retrieval and ranking performance. These results support the effectiveness of preserving task-specific capacity while maintaining task-shared knowledge. 
Meanwhile, removing uncertainty estimation also leads to performance degradation on both tasks, indicating the effectiveness of assigning relatively higher weights to gradients with more stable historical directions when gradient conflict occurs.
Moreover, we also observe that the removal of knowledge distillation mainly degrades retrieval performance. This observation demonstrate the crucial role of knowledge distillation to transfer knowledge from ranking to retrieval.

        

\subsection{Gradient Conflict Analysis (RQ3)}

To investigate whether LazUnirec alleviates gradient conflict, we compare the ordinary conflict (\textit{i.e.}, cos < 0) and robust conflict (\textit{i.e.}, cos < -0.05) percentages within the Attention and FFN components against the vanilla Transformer in Figure~\ref{gradient_conflict_analysis}. 
Compared with the vanilla Transformer, LazUnirec achieves relative reductions of 9.16\% and 13.23\% in the ordinary and robust conflict percentages within Attention, as well as corresponding reductions of 14.32\% and 37.76\% within FFN.
These observations demonstrate the effectiveness of the task-shared and task-private paradigm. Specifically, LoRA-modulated attention provides task-private adaptations to accommodate different objectives while preserving shared attention parameters. Meanwhile, task-adaptive sparse MoE assigns dedicated experts to retrieval and ranking and selectively activates shared experts, alleviating gradient conflict while maintaining knowledge sharing. The particularly pronounced reduction in robust conflict further suggests that LazUnirec effectively alleviates stronger gradient conflict during joint optimization.

\begin{figure}[t]
\setlength{\abovecaptionskip}{0.cm}
	\begin{center}
        \subfigure
        {\begin{minipage}[b]{.49\linewidth}
        \centering
        \includegraphics[scale=0.20]{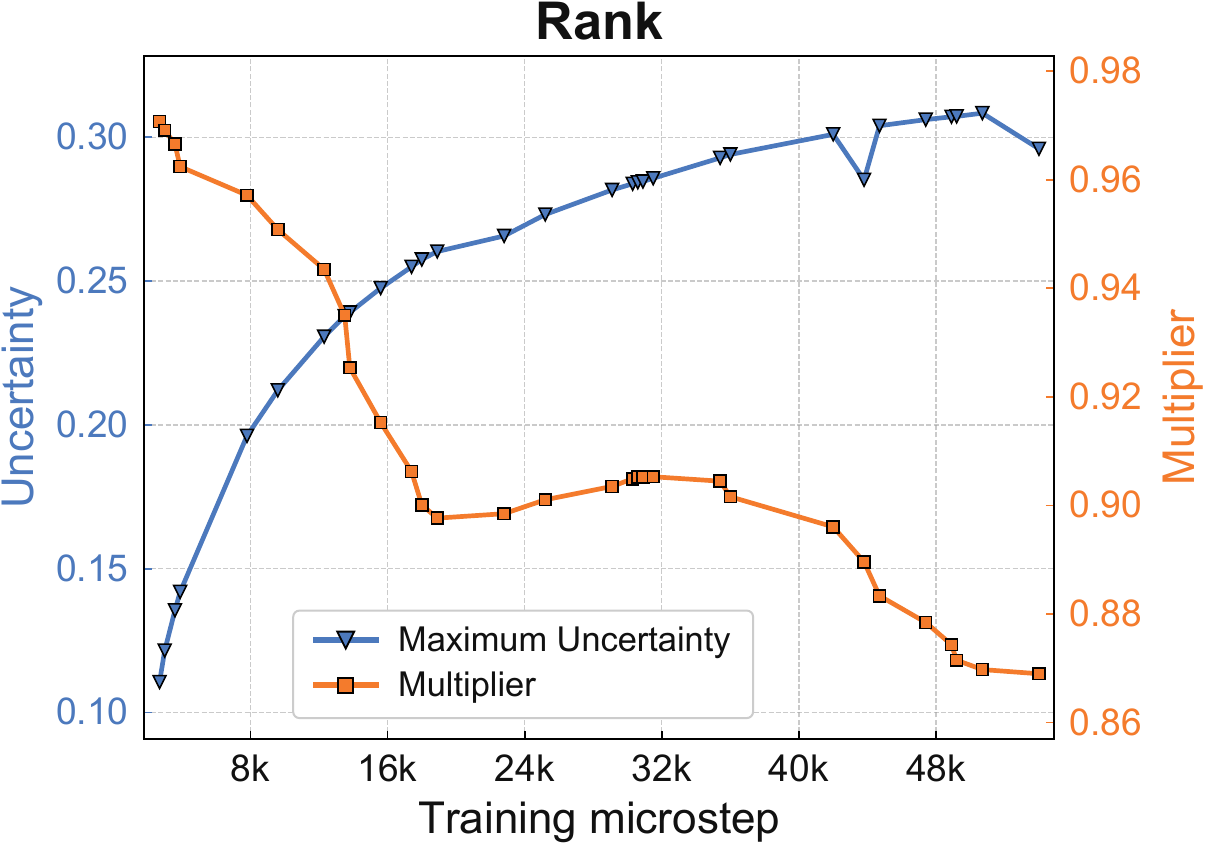}
        \end{minipage}}
        \subfigure
        {\begin{minipage}[b]{.49\linewidth}
        \centering
        \includegraphics[scale=0.20]{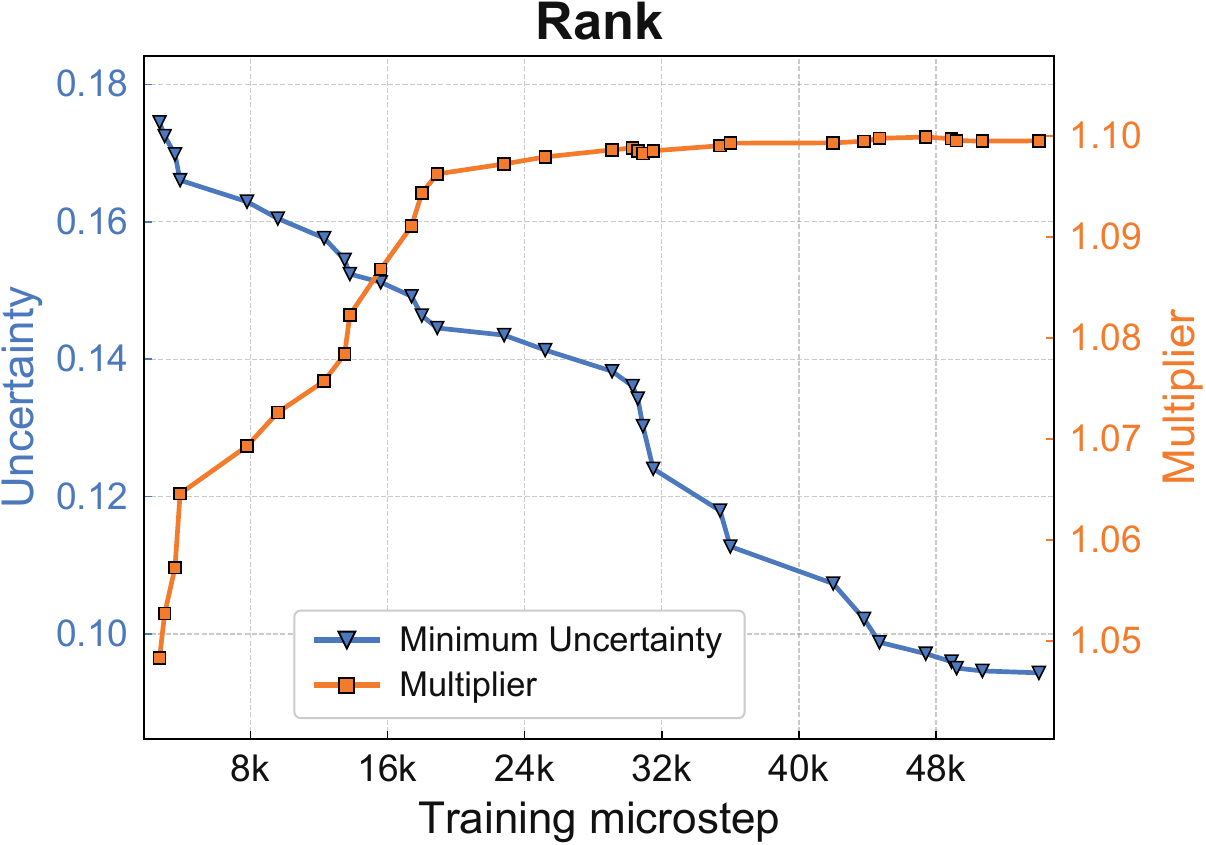}
        \end{minipage}}
        \subfigure
        {\begin{minipage}[b]{.49\linewidth}
        \centering
        \includegraphics[scale=0.20]{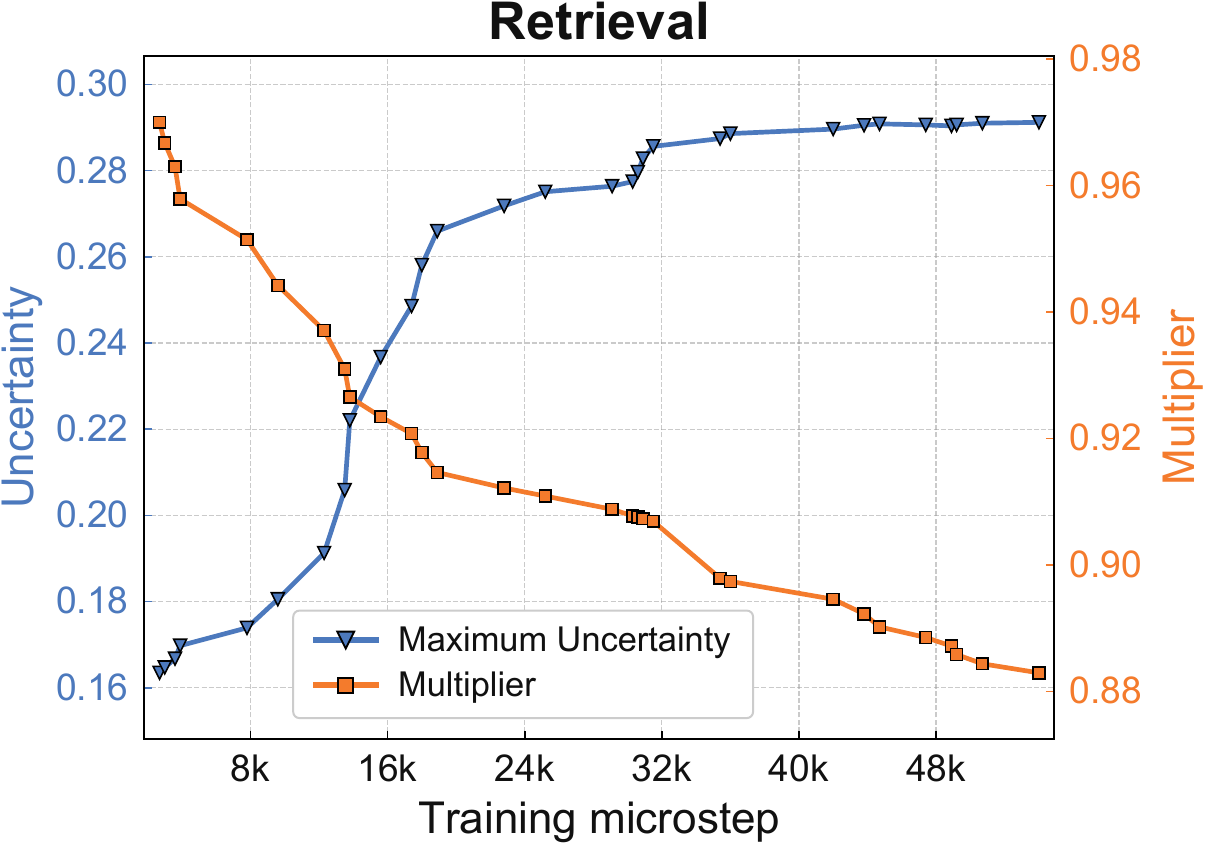}
        \end{minipage}}
        \subfigure
        {\begin{minipage}[b]{.49\linewidth}
        \centering
        \includegraphics[scale=0.20]{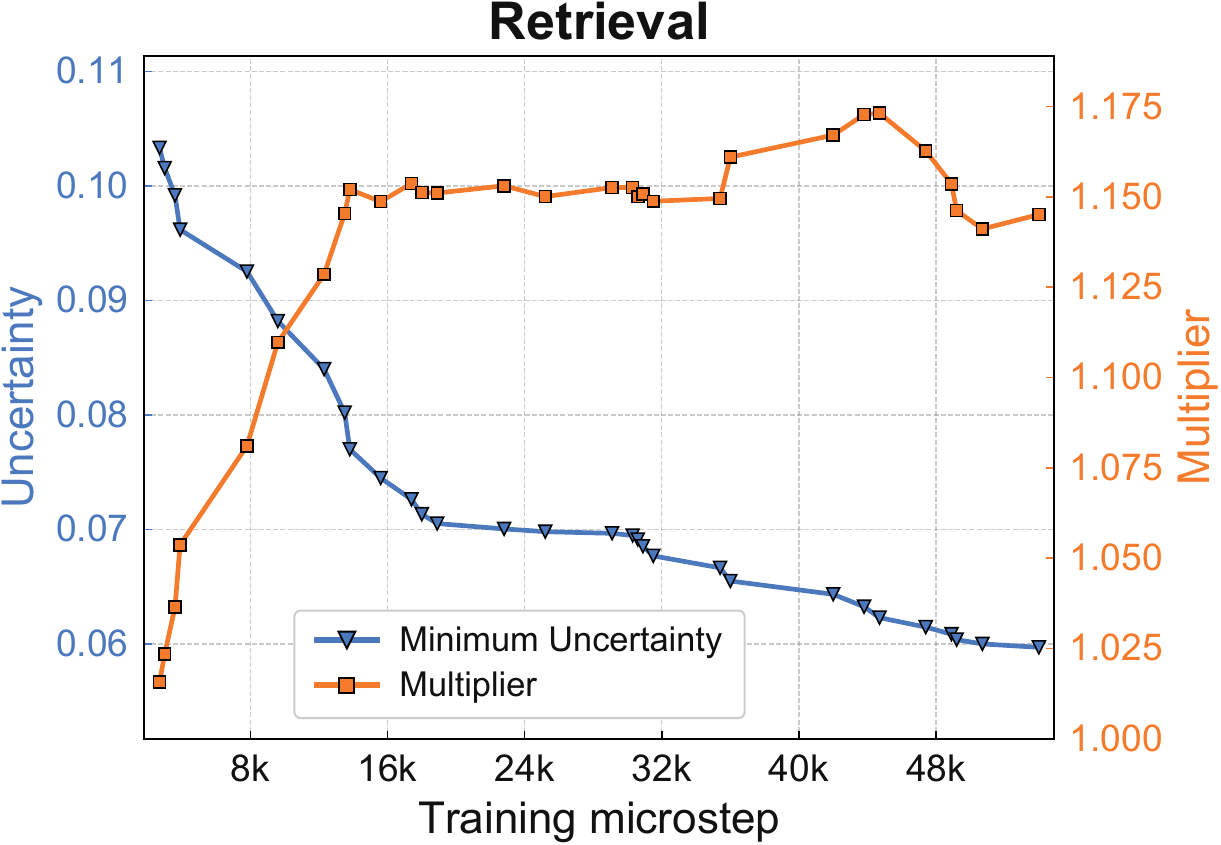}
        \end{minipage}}
        
	\caption{Relative down-weighting/up-weighting multipliers corresponding to sample points with maximum/minimum uncertainty during retrieval and ranking.}
	\label{uncertainty_estimation_analysis}
	\end{center}
\end{figure}

\subsection{Uncertainty Estimation Analysis (RQ4)}

To investigate how uncertainty estimation guides the direction of gradient updates, we randomly sample 30 maximum uncertainty observations and 30 minimum uncertainty observations during training for retrieval and ranking, respectively. For each observation, we record the uncertainty $\mathcal{E}_s^\tau$ and its corresponding loss-weighting multiplier $\omega_s^\tau$, as shown in Figure~\ref{uncertainty_estimation_analysis}.

From the results, we observe that high uncertainty observations receive lower multipliers, reducing their relative contributions to the total losses. As their uncertainty generally increases during training, the corresponding multipliers tend to decrease. Meanwhile, low uncertainty observations receive higher multipliers, increasing their relative contributions. This trend is consistently observed in both retrieval and ranking, indicating that uncertainty estimation adaptively adjusts the influence of different observations according to gradient reliability.
Since uncertainty reflects the historical fluctuation of gradient directions, down-weighting high uncertainty observations reduces the influence of unstable gradients, while up-weighting low uncertainty observations emphasizes more stable gradients. Consequently, the parameter updates place greater emphasis on reliable directions when gradient conflict occurs.

\begin{table}[t]
\centering
\caption{Scenario ablation study on Scenarios 1, 2, and X.}
\label{exp_scenario_ablation}
\setlength\tabcolsep{2.0pt}
\begin{tabular*}{0.49\textwidth}
{@{\extracolsep{\fill}}@{}lcccccc@{}}
\toprule
\multirow{3.5}{*}{\textbf{Variants}} &
\multicolumn{2}{c}{\textbf{Scenario 1}} &
\multicolumn{2}{c}{\textbf{Scenario 2}} &
\multicolumn{2}{c}{\textbf{Scenario X}} \\
\cmidrule(lr){2-3}\cmidrule(lr){4-5}\cmidrule(lr){6-7}
&
\textbf{CTR} & \textbf{HR} &
\textbf{CTR} & \textbf{HR} &
\textbf{CTR} & \textbf{HR} \\
\cmidrule(lr){2-2}\cmidrule(lr){3-3}
\cmidrule(lr){4-4}\cmidrule(lr){5-5}
\cmidrule(lr){6-6}\cmidrule(lr){7-7}
&
\textbf{GAUC} & \textbf{@100} &
\textbf{GAUC} & \textbf{@100} &
\textbf{GAUC} & \textbf{@100} \\
\midrule
\textit{w/} Scenario 1 & 0.6647 & 0.3560 & N.A. & N.A. & N.A. & N.A. \\
\textit{w/} Scenario 2 & N.A. & N.A. & 0.6777 & 0.5706 & N.A. & N.A. \\
\textit{w/} Scenario X & N.A. & N.A. & N.A. & N.A. & 0.6417 & 0.3086 \\
LazUnirec              & \textbf{0.6722} & \textbf{0.3620} & \textbf{0.6828} & \textbf{0.5906} & \textbf{0.6572} & \textbf{0.3686} \\
\bottomrule
\end{tabular*}
\end{table}

\subsection{Multi-scenario Analysis (RQ5)}
\label{multi_scenario_analysis}

In this section,
we select trigger-free Scenario 1, trigger-based Scenario 2, and trigger-free long-tail Scenario X from over a dozen scenario in total to evaluate the effectiveness of multi-scenario modeling. For each selected scenario, we construct a single-scenario variant by retaining only its training data and excluding all other scenarios. We compare these variants with LazUnirec in terms of CTR GAUC and HR@100, as shown in Table~\ref{exp_scenario_ablation}.

From the results, we find that multi-scenario training improves both retrieval and ranking across all three scenarios. This suggests that interactions from different scenarios provide complementary supervision, enabling LazUnirec to learn transferable user preferences through shared parameters. 
Besides, the improvements are particularly pronounced in long-tail Scenario X, indicating that supervision from other scenarios can help alleviate data sparsity in the target scenario. 

\begin{table}[t]
\centering
\caption{Online A/B testing results.}
\label{tab:online_AB}
\setlength{\tabcolsep}{10pt}
\begin{tabular*}{0.49\textwidth}
{@{\extracolsep{\fill}}@{}lcccc@{}}
\toprule
\textbf{Scenarios} & \textbf{IPV} &  \textbf{Orders} & \textbf{Buyers} & \textbf{GMV} \\
\midrule
Trigger-free & +5.25\%  & +5.75\% & +4.57\% & +4.01\% \\
\midrule
Trigger-based & +4.01\%  & +4.64\% & +3.06\% & +3.06\% \\
\bottomrule
\end{tabular*}
\end{table}

\subsection{Online Efficiency Analysis (RQ6)}

To evaluate the online efficiency of LazUnirec, we investigate the online QPS gains from the four deployment optimizations.
Specifically, KV cache improves online QPS by 20\% by reusing user sequence keys and values across retrieval and ranking, as well as across requests.
Kernel fusion increases online QPS by 44\% by reducing memory traffic and kernel launch overhead.
Moreover, low-precision serving achieves a 5\% improvement in online QPS through selective FP8 computation within SwiGLU.
Cross-scenario batching further improves online QPS by 59\% by aggregating requests from different scenarios into larger batches, thereby improving GPU utilization.

\subsection{Online A/B Testing (RQ7)}

We deploy LazUnirec in both trigger-free and trigger-based scenarios on a large-scale industrial e-commerce platform. LazUnirec simultaneously replaces the original MIMA-based retrieval~\cite{mima} and LazFormer-based ranking~\cite{lazformer}. 
The online A/B testing was launched in October 2026, last for one week, with 10\% online traffic. The results are shown in Table~\ref{tab:online_AB}.
Specifically, in trigger-free scenarios, LazUnirec achieves relative improvements of 5.25\% in IPV (\textit{i.e.}, item page views), 5.75\% in orders, 4.57\% in buyers, and 4.01\% in GMV. In trigger-based scenarios, the corresponding improvements are 4.01\%, 4.64\%, 3.06\%, and 3.06\%, respectively.
The online results show that LazUnirec improves user engagement and revenue, supporting its effectiveness and business value in real-world multi-scenario industrial recommendation.

\section{Related Work}


Transformer~\cite{trans} has demonstrated promising performance in LLMs, while recent ranking models~\cite{onetrans,mixformer,hyformer,sort,lazformer,hstu,mtgr} have investigated \textbf{Transformer for industrial recommendation}. Specifically, OneTrans~\cite{onetrans} integrates user sequence modeling and feature interaction within a shared Transformer backbone. MixFormer~\cite{mixformer} enables joint scaling of dense feature interaction and sequence modeling, while HyFormer~\cite{hyformer} iteratively combines long-sequence decoding with heterogeneous feature interaction. 
Moreover, LazFormer~\cite{lazformer} investigates transferable generative pre-training to alleviate the negative transfer of dense parameters and the overfitting of sparse parameters.
These studies demonstrate the potential of Transformers for scalable ranking through improved architectures and training strategies. 
However, these advances mainly benefit the ranking stage, while the architectural similarity between Transformer-based ranking and retrieval provides a foundation for integrating the two stages within a unified model.
Building on this direction, LazUnirec applies Transformer to share user sequence encoding across retrieval and ranking, and further addressing the gradient conflict arising from their different optimization objectives.


\textbf{Unifying retrieval and ranking} has attracted increasing attention~\cite{unipinrec,oneranker,unir2,unisgr,onetrans_v2,onepiece} for reducing redundant user sequence encoding and facilitating cross-task knowledge sharing. UniPinRec~\cite{unipinrec} unifies input formats, training, and serving through a shared Transformer with task-specific heads. 
$\operatorname{UniR^2}$~\cite{unir2} organizes user context, semantic ID trajectories, and item features within a single decoder sequence, introducing ranking-specific LoRA and gradient isolation to control optimization interference. UniSGR~\cite{unisgr} combines multi-scenario pre-training with joint generation and multi-objective ranking, using task-aware tokens to align the two tasks. Furthermore, OneTrans-V2~\cite{onetrans_v2} extends unification to retrieval, pre-ranking, and fine-ranking, sharing user sequence computation and scaling the backbone with sparse MoE.
However, these methods overlook the coordination of task-specific adaptation and reliable shared-parameter updates across scenarios, which leads to gradient conflict issue.
LazUnirec allows both retrieval and ranking to optimize shared parameters, while introducing task-private capacity into Attention and FFN components to alleviate gradient conflict. Moreover, it estimates gradient reliability from historical directional stability to guide joint optimization, and incorporates knowledge distillation for explicit cross-task knowledge transfer.

\vspace{-0.2cm}
\section{Conclusion}

In this paper, we propose LazUnirec, a unified retrieval and ranking approach for multi-scenario industrial recommendation. We first introduce LoRA-modulated attention and task-adaptive sparse MoE to alleviate gradient conflict on the Attention and FFN components by constructing task-shared and task-private parameters. Additionally, we present uncertainty estimation to evaluate gradient reliability of retrieval and ranking, thus guiding the model toward a more reliable gradient update direction when conflict occurs. Moreover, the knowledge distillation further establishes an explicit knowledge transfer path from ranking to retrieval. Extensive offline experiments and online A/B testing demonstrate the effectiveness of LazUnirec. LazUnirec has been deployed in a real-world multi-scenario recommendation system, delivering significant gains for the platform.

\balance
\bibliographystyle{ACM-Reference-Format}
\bibliography{sample-base-extend.bib}

\end{document}